\documentclass[aps,prb,twocolumn,groupedaddress,showpacs]{revtex4}
\usepackage[dvips]{color}
\usepackage{graphicx}% Include figure files

\begin{document}

%\preprint{}

\title{Three-terminal scanning tunneling spectroscopy of suspended carbon nanotubes}


\author{B.J. LeRoy}
\author{J. Kong}
\author{V.K. Pahilwani}
\author{C. Dekker}
\author{S.G. Lemay}
\email[]{lemay@mb.tn.tudelft.nl}
%\thanks{}
\affiliation{Kavli Institute of Nanoscience, Lorentzweg 1, 2628 CJ
Delft, The Netherlands}

\date{\today}

\begin{abstract}
We have performed low-temperature scanning tunneling spectroscopy
measurements on suspended single-wall carbon nanotubes with a gate
electrode allowing three-terminal spectroscopy measurements. These
measurements show well-defined Coulomb diamonds as well as side
peaks from phonon-assisted tunneling.  The side peaks have the
same gate voltage dependence as the main Coulomb peaks, directly
proving that they are excitations of these states.
\end{abstract}


\maketitle

%\section{\label{Introduction}Introduction}
The interplay between electrical and mechanical degrees of freedom
is critical for understanding single-molecule devices.  Changes in
molecular orientation and electron-phonon coupling can have a
large effect on the transport through the molecule.  While there
has been much theory\cite{Boese, Fedorets, Alexandrov, McCarthy,
Flensberg, Aji, Mitra} on this interplay, there have been
relatively few experiments to date\cite{Park, Yu, Pasupathy}. The
primary difficulty in the latter has been making reproducible
measurements due to variations in how the molecule couples to the
electrodes. Present state-of-the-art techniques require
fabricating many devices of which only a small percentage show the
desired results. For fundamental studies, better control can be
achieved using a scanning tunneling microscope (STM) tip as one of
the contacts and performing combined scanning tunneling
spectroscopy and electrical transport measurements. This allows
the position and coupling of the electrode to be varied.

Single-walled carbon nanotubes (SWCNTs) are ideal single-molecules
because of their well understood electrical properties and their
long length that allows spatially resolved measurements. By
suspending the nanotubes, the coupling to the metallic substrate
is reduced, revealing Coulomb blockade behavior\cite{LeRoyAPL}. In
the center of the suspended regions, we have observed additional
side peaks on the Coulomb peaks which we have previously
attributed to phonon-assisted tunneling \cite{LeRoyNature}.  The
assignment of the additional peaks as excitations of the main
Coulomb peaks was due to three experimental observations.  The
peaks were equally spaced in voltage from the main peaks.
Additional peaks appeared at higher current in analogy to
photon-assisted tunneling \cite{Tien}. Lastly, the peak energy
scaled as the inverse of the SWCNT diameter in agreement with the
energy of the radial breathing mode phonon. However, the lack of a
gate electrode precluded a direct confirmation that the peaks
where excitations of the main Coulomb peak.

In this paper, we show results from three-terminal scanning
tunneling spectroscopy measurements on suspended SWCNTs.  We have
added a gate electrode into the device design in order to allow
three-terminal electrical measurements.  Using the gate electrode,
we are able to access the full Coulomb diamond as a function of
both substrate and gate voltage and to accurately determine the
capacitances between the SWCNT and the tip, substrate, and gate.
We are also able to determine the origin of the phonon-assisted
tunneling side peaks, confirming that the peaks are indeed
excitations of the main elastic Coulomb peaks. Furthermore, we
report additional features in the spectra, in particular avoided
crossings between sets of differential conductance peaks.

%\section{\label{Sample Preparation}Sample Preparation}
The samples consisted of individual single-wall carbon nanotubes
freely suspended across trenches etched in SiO$_2$ or Si$_3$N$_4$.
The sample preparation has been described previously for the
Si$_3$N$_4$ samples \cite{KongAPL}.  The fabrication of the
SiO$_2$ samples was similar except that the trenches were only
etched 150 nm into the 250 nm thick oxide layer.  This was
followed by deposition of Pt at a 60 degree angle from the normal
to prevent the metal from reaching the bottom of the trench. With
this method, we can use the Si as a back gate because there is no
metal in the trench to screen the gate.  This also eliminates the
possibility of short circuits between the Pt substrate and the
gate electrode because the gate is still covered by 100 nm of
SiO$_2$.  During scanning tunneling microscopy imaging of the
Si$_3$N$_4$, we observed that SWCNTs could be pushed around the
surface whereas they were more firmly attached on the SiO$_2$
sample. The reason for this difference is not understood.  The
only difference that we observed in the electrical properties of
the two sets of samples is that the gate electrode coupled better
to the SWCNTs in the Si$_3$N$_4$ sample because of its closer
proximity to the SWCNTs.

\begin{figure}
\includegraphics[width=3.35in]{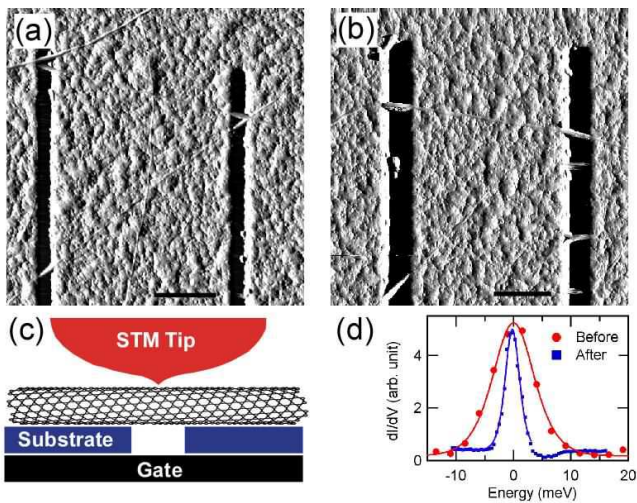}%
\caption{(color online) AFM images of SWCNTs crossing (a) 100 and
(b) 200 nm wide trenches. The scale bars are 500 nm.  (c)
Schematic diagram showing the setup used to perform spectroscopy
on suspended SWCNTs.  The STM tip acts as the source while the
substrate is the drain.  The gate electrode is either formed by Pt
at the bottom of the trench or the Si substrate. (d) Differential
conductance as a function of energy before (red circles) and after
(blue squares) heat sinking the wires in the STM. The solid lines
are fits of the derivative of the Fermi distribution.}
\label{setup}
\end{figure}

Figures \ref{setup}(a) and (b) are atomic force microscope (AFM)
images of SWCNTs grown across 100 and 200 nm wide trenches,
respectively. Figure \ref{setup}(c) is a schematic diagram showing
a SWCNT crossing a trench with the STM tip, substrate and gate
labelled.

All the STM measurements were made in a low-temperature ultra-high
vacuum scanning tunneling microscope (STM) operating at 5 K.  The
commercial STM (Omicron LT-STM) has been modified to decrease the
electron temperature and allow three-terminal measurements. In
order to decrease the electron temperature, all the wires going to
the STM were heat sunk to a large Cu rod connected to the Li-He
vessel. This lowered the effective electron temperature from 25 K
to about 5 K. This can be seen in Fig. \ref{setup}(d) where two
Coulomb blockade peaks are plotted before (red) and after (blue)
the modifications to the STM. Although the electron temperature
was greatly reduced the temperature measured by the Si sensor
diode remained nearly constant at 4.6 K. We have also modified the
sample holders to allow three electrical contacts to be made. This
allows separate source, drain, and gate contact to the sample.

SWCNTs crossing trenches were located using scanning tunneling
microscopy. Typical settings for STM measurements were a gap
voltage of -0.5 V applied to the Pt substrate with respect to the
grounded STM tip, and a feedback current of 300 pA. After finding
suspended SWCNTs, spectroscopy measurements were performed on the
suspended portion. The spectroscopy was performed by stabilizing
the current and then turning off the feedback circuit and
measuring the current as a function of the substrate voltage. The
differential conductance, dI/dV was measured by adding a small ac
voltage (2-4 mV rms, 887 Hz) and using lock-in detection.

\begin{figure}
\includegraphics[width=3.35in]{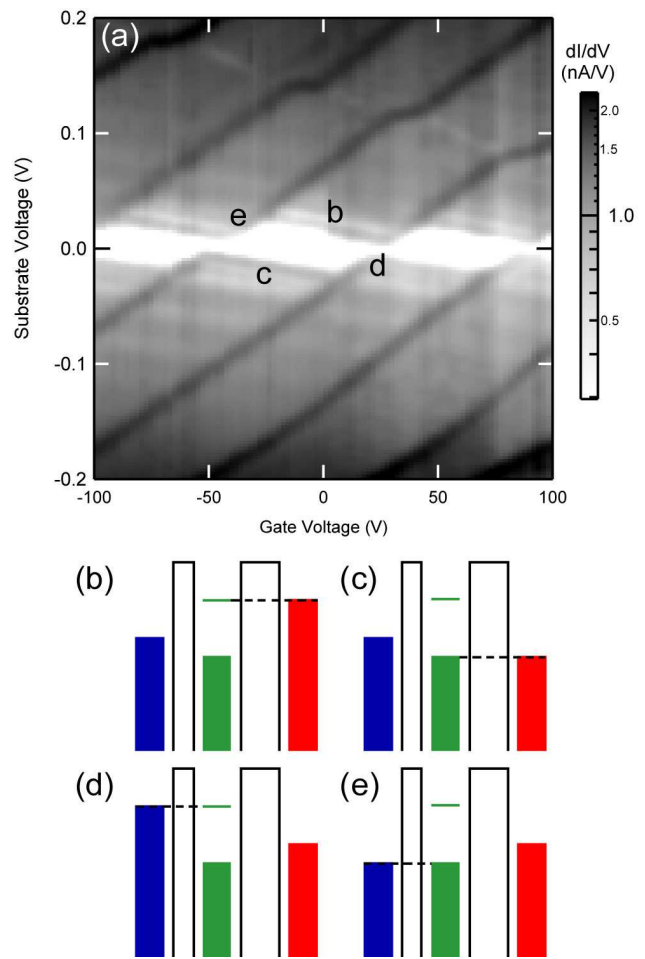}%
\caption{(color online) (a) Log of differential conductance as a
function of gate and sample voltage, showing Coulomb diamonds. The
data is on a SWCNT, which is suspended over a 100 nm wide trench.
The feedback current was set to 300 pA at -0.2 V. (b)-(d) Diagrams
showing from where the edges of the diamonds arise.  The strong
lines, (d) and (e) come from the Fermi level in the substrate
lining up with a state on the SWCNT. On the other hand, the
fainter lines (b) and (c) come from a state in the SWCNT lining up
with the Fermi level of the tip. } \label{diamonds}
\end{figure}

%\section{\label{Gate Effect}Gate Effect}
The addition of a gate electrode in the scanning tunneling
microscopy setup allows the electronic behavior of the suspended
nanotubes to be fully investigated.  By changing the gate voltage,
the number of electrons on the nanotube changes. Figure
\ref{diamonds}(a) plots the differential conductance as a function
of gate and substrate voltage. Diamond-shaped regions of zero
current (white area) are visible, demonstrating the Coulomb
blockade effect\cite{Grabert,Kastner}. When the substrate voltage
is zero, moving horizontally from one diamond to the next
corresponds to changing the number of electrons on the SWCNT by
one.  The edges of the diamond correspond to the Fermi level of
either the tip or the substrate lining up with a filled or empty
state of the SWCNT.

Access to the full Coulomb diamond plot allows the unambiguous
identification of the origin of all the peaks in the measurements.
The strong peaks running from the bottom left to the top right of
Fig. \ref{diamonds}(a) are due to the Fermi level of the substrate
aligning with a state on the SWCNT. A schematic of the energy
levels for these lines are shown in Fig. \ref{diamonds} (d) and
(e). These lines are strong because of the large asymmetry in the
tunnel barriers.  When the Fermi energy of the substrate passes an
empty state on the dot, the state is filled and the energy of the
other states shift upward by the charging energy. This brings many
more states into the bias window which can then tunnel out the tip
lead giving a large peak in the conductance. The weaker lines
running in the other direction are from the tip aligning with
states in the SWCNT. Fig. \ref{diamonds} (b) and (c) are
schematics showing the energy levels giving rise to these lines.
The lines appear much weaker because the total tunnel rate through
the SWCNT is controlled by the tip lead and this process only
leads to one new state that electrons from the tip can tunnel
into.  Therefore there is only a small change in the current and
the peaks are only visible at low sample voltage and current.

From the size of the Coulomb diamonds and the slope of their
edges, we can determine the three capacitances in our measurement
as we have done previously\cite{KongAPL}. The three capacitances
are the tip to the SWCNT, C$_{tip}$, the SWCNT to the substrate,
C$_{sub}$, and the SWCNT to gate capacitance, C$_g$.  The gate
capacitance is found using the equation $C_g = e/V_g$ where V$_g$
is the width of the diamond in gate voltage. This equation assumes
that the level spacing of the SWCNT is small.  If this is not the
case, then the widths of the Coulomb diamonds would vary because
of the additional energy caused by the level spacing. For this
SWCNT, we see no variation in the size of the Coulomb diamonds and
therefore we conclude that the level spacing is small.  From the
width of the diamond, we find a value of C$_g$ of 2 zF. The height
of the diamond gives the charging energy E$_c$. This is related to
the total capacitance C$_\Sigma$ using the equation $E_c = e /
C_\Sigma$. From the height of the Coulomb diamond, we get
C$_\Sigma$ $\approx$ 6 aF. The slope of the edge of the diamond
going towards the bottom right is given by $-C_g/C_{sub}$ allowing
us to determine the value of C$_{sub}$. We find that C$_{sub}$
$\approx$ 4.6 aF. The last capacitance can be found by the slope
of the other lines, $C_g/(C_g + C_{tip})$, yielding C$_{tip}$
$\approx$ 1.4 aF.  As a further consistency check, the sum of the
three capacitances can be compared with C$_\Sigma$.

% Comparison to expectations
The experimental values found for the capacitances are in
reasonable agreement with electrostatic simulations of our device
structure. We modelled the tip as a $\sim$150 nm radius sphere and
the SWCNT as a 2 nm diameter cylinder.  We then calculated the
capacitance between the SWCNT and the tip, substrate and gate
electrode.  We find that the calculated C$_{sub}$ is about 2 aF
but depends on the length of the SWCNT quantum dot over the Pt.  A
simple estimate for the capacitance of the SWCNT to the substrate
can be given by considering an infinite cylinder lying above a
conducting plane. This has a capacitance per unit length of $2 \pi
\epsilon_0 /\ln[(d+(d^2-R^2)^{1/2})/R]$ where d is the distance of
the center of the cylinder above the plane and R is the radius
\cite{Scott}. If we use d = 1.25 nm and R = 1.0 nm, then the
capacitance is 2.5 aF for every 30 nm. From this, we conclude that
our SWCNT quantum dot does not extend very far over the Pt
substrate.

In previous measurements on suspended SWCNTs, we have observed
additional peaks which were attributed to phonon-assisted
tunneling. \cite{LeRoyNature}  The assignment of the peaks as
phonon side peaks was based mainly on three factors: they were
equally spaced from the main Coulomb peaks, additional peaks
appeared at high current, and their energy varied as a function of
SWCNT diameter. Despite this evidence, another possibility was
that the side peaks were not excitations of the main Coulomb peak
but rather due to tunneling at the other junction. In this
scenario, the appearance of new peaks at high currents would be
the result of a decreasing $R_{tip}$.  While this scenario was
found to be unlikely based on quantitative analysis of peak
spacing, a gate electrode provides more definitive evidence; if
the peaks were due to the tip lead, the sign of their slope with
gate voltage would be the opposite of the main Coulomb peaks.

\begin{figure}
\includegraphics[]{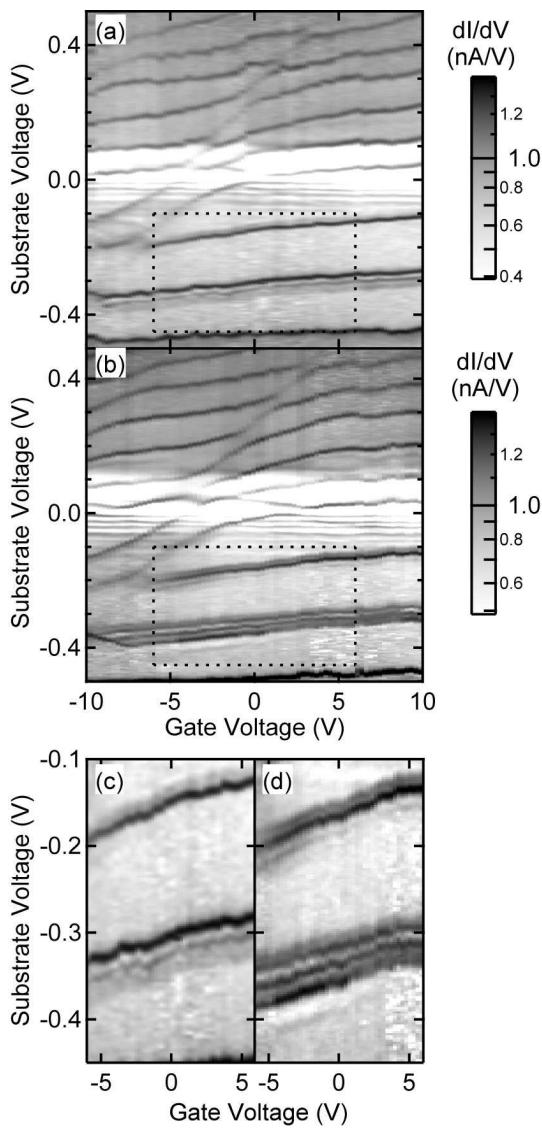}%
\caption{Log of differential conductance as a function of gate and
substrate voltage at two different setpoint currents, (a) 300 pA
at -0.5 V and (b) 500 pA at -0.5 V. Side peaks due to
phonon-assisted tunneling are visible in the negative substrate
voltage region. (c) and (d) Zoom-in on the boxes in (a) and (b)
respectively.  The side peaks due to phonon-assisted tunneling are
running parallel to the main Coulomb peaks, demonstrating that
they are excitations of this state. This SWCNT is suspended across
a 200 nm wide trench.} \label{phonons}
\end{figure}

Figure \ref{phonons} shows the differential conductance as a
function of sample and gate voltage at two different setpoint
currents. Fig. \ref{phonons}(a) is taken with the feedback current
stabilized at 300 pA at -0.5 V while Fig. \ref{phonons}(b) has a
current of 500 pA at -0.5 V.  Once again, there are peaks running
from the bottom left to the top right due to the addition of
electrons to the SWCNT. These peaks are due to the Fermi level in
the substrate lining up with a state on the SWCNT. The peaks
running in the other direction, due to the Fermi level of the tip
lining up with a state on the SWCNT, are too faint to be resolved
in this measurement due to the large tip resistance. However,
there are side peaks running parallel to the main peaks due to
phonon-assisted tunneling \cite{LeRoyNature}. In this case, the
side peaks are most prominent for negative sample voltages. Figs.
\ref{phonons}(c) and (d) are zoom-ins on the negative voltage
region of (a) and (b) respectively.  The fact that the side peaks
run parallel to the main peak confirms that they are an excitation
of this state, ruling out the possibility that they are the faint
peaks due to the Fermi level of the tip lining up with a state on
the SWCNT. Therefore, we can assign these phonon peaks to
electrons tunneling between the SWCNT and the substrate lead.
Knowing the origin of the phonon side peaks allows their energy to
be calculated from their spacing from the Coulomb peak. Since they
are due to the substrate lead, we must scale their measured
voltage from the Coulomb peak by the fraction of the voltage that
drops at this junction.  This gives a scaling factor of
$C_{tip}/C_{\Sigma}$. The incorporation of the gate electrode and
measurement of the Coulomb diamond allows a more accurate
determination of the phonon energy because the capacitances can be
better measured from the slopes of the diamonds.

\begin{figure}
\includegraphics[width=3.35in]{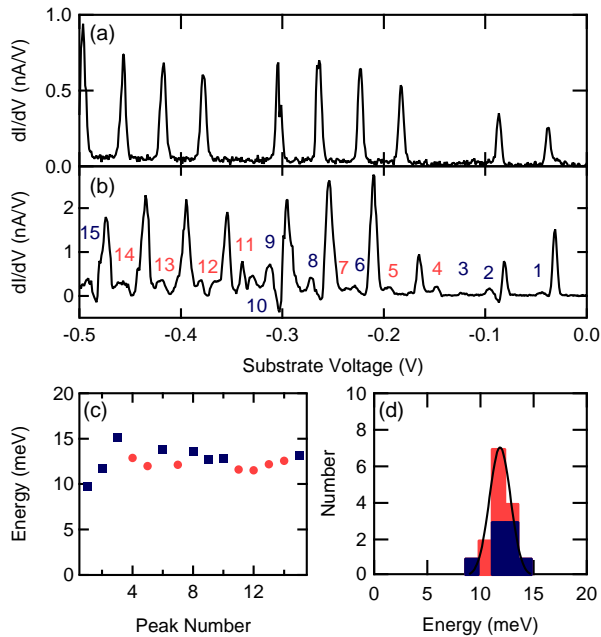}%
\caption{(color online) (a) Differential conductance as a function
of substrate voltage showing Coulomb peaks due to the addition of
electrons to the SWCNT.  The feedback current was stabilized at
100 pA at -0.6 V.  (b) Same as (a) but the feedback current was
set to 300 pA. Now additional side peaks are visible due to
phonon-assisted tunneling.  The side peaks are labelled by blue
and red numbers corresponding to emission and absorption.  (c)
Energy of side peaks as a function of peak number. Blue squares
correspond to emission peaks while red circles are for absorption
peaks (d) Histogram of side peak energy for both emission and
absorption showing that all of the side peaks have similar energy.
The solid line is a gaussian fit to the histogram.}
\label{spacings}
\end{figure}

%\section{\label{Origin of Phonons}Origin of Phonons}
The strength of the phonon side peaks can be controlled by the
current through the SWCNT \cite{LeRoyNature}. Figs.
\ref{phonons}(c) and (d) demonstrate the effect of increasing the
current.  At higher current, \ref{phonons}(d), the side peaks are
more pronounced and additional ones are present.  Figures
\ref{spacings}(a) and (b) plot the differential conductance as a
function of energy at two different setpoint currents.  At low
setpoint current, 100 pA at -0.6 V (Fig. \ref{spacings}(a)),
groups of four Coulomb peaks are visible but there are no phonon
side peaks. The groupings are caused by the two spin-degenerate
bands of the SWCNT. This allows level spacing in this SWCNT to be
measured, which we find to be about 30 meV. At higher setpoint
current, 300 pA (Fig. \ref{spacings}(b)), additional peaks become
visible. The side peaks are equally spaced in energy from their
associated Coulomb peaks and can occur on either side of their
associated Coulomb peak. This implies that they are due to both
emission and absorption of phonons. The fact that they are equally
spaced can be seen in Figure \ref{spacings}(c), which plots the
peak energy for a series of 15 side peaks.  The blue squares plot
the energy for peaks associated with emission of a phonon, while
the red circles are for absorption.  These energies have been
converted from the substrate voltage using the scaling factor
$C_{tip}/C_{\Sigma}$. This takes into account that only a fraction
of the voltage drops at the substrate-SWCNT junction. Because this
device did not have a gate electrode, the values of $C_{tip}$ and
$C_{sub}$ were determined from the spacing of the Coulomb peaks
and the values of the differential conductance \cite{Hanna}.
Figure \ref{spacings}(d) is a histogram of the energies showing
that both the absorption and emission side peaks have similar
energies. For this 2.5 nm tube the phonon energy is measured to be
11.8 $\pm$ 1.4 meV. The uncertainty corresponds to the standard
deviation of the Gaussian fit in Fig. \ref{spacings}(d). This
agrees well with the theoretical energy of the radial breathing
mode, which is 11.5 meV. \cite{Dresselhaus}

The SWCNT phonon density of states includes many more modes which
are not observed in the measurements.  The low-energy string modes
can not be resolved because their energy is much lower than kT.
For a 100 nm long SWCNT, their energy is 0.014 meV which is 30
times smaller than kT.\cite{Sapmaz} Recent calculations show that
the electron-phonon matrix element is largest near $k$=0 for the
radial-breathing and optical phonons \cite{Perebeinos}. However,
experimentally the high-energy optical phonons ($\approx$200 meV)
are not observed. The theoretical calculations do not consider the
local tunneling nature of the STM experiment, which may influence
the relative strengths of the phonon peaks. In particular, the
speed that the electron spreads out around the SWCNT compared to
the frequency of the phonon makes circumferentially symmetric
modes more likely to be excited. The high energy, long wavelength
and high symmetry of the radial-breathing mode make it the easiest
phonon to excite.

\begin{figure}
\includegraphics[width=3.35in]{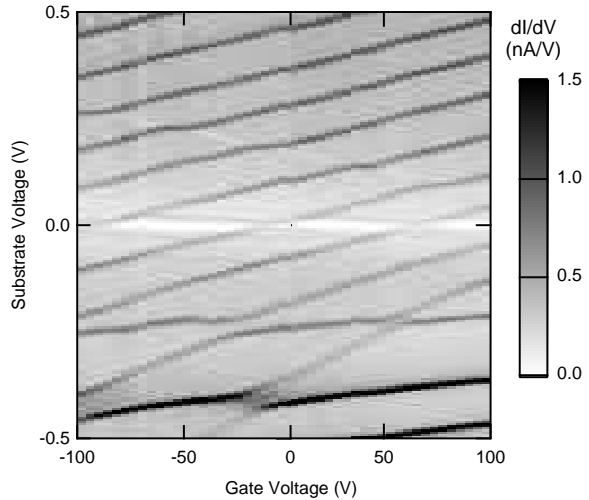}%
\caption{Differential conductance as a function of gate and
substrate voltage, showing peaks due to the addition of electrons.
For negative substrate voltages, there are sets of peaks that show
an avoided crossing behavior.  Data is taken at the suspended
portion of a metallic SWCNT. The tube is suspended over a distance
of 100 nm.  Feedback current is set to 300 pA at -0.5 V.}
\label{crossings}
\end{figure}

%\section{\label{Avoided Crossings}Avoided Crossings}
In most of our measurements of differential conductance as a
function of gate and sample voltage, we observe some Coulomb peaks
that do not run parallel to the others.  Figure \ref{crossings}
shows an example of this in the negative substrate-voltage portion
of the plot.  There are two peaks that are only weakly dependent
on the gate voltage, being nearly horizontally.  This implies that
these peaks have a different coupling to the gate from the main
series of peaks. There is also a faint dip in the positive
substrate voltage region whose origin is unknown. Because not all
the peaks are parallel, there are locations where they cross.
Three such crossings can be seen in Fig. \ref{crossings}. When
this occurs, the peaks show an avoided crossing behavior rather
than a simple intersection. The fact that the peaks do not cross
implies that there is a coupling between these two states. The
origin of these extra peaks is unknown, but this type of feature
may arise from localized states caused by a defect in the SWCNT,
or a second quantum dot in parallel. Based on topography and
spectroscopy, we believe that all of the nanotubes that we have
measured have been isolated singe-walled tubes. Therefore, it is
unlikely that the avoided crossings arise from interaction between
two SWCNTs in a rope. Future measurements probing the spatial
extent of these extra peaks may allow their origin to be
identified.

%\section{\label{Conclusions}Conclusions}
In summary, we have shown scanning tunneling spectroscopy
measurements on the free-standing portion of suspended SWCNTs with
an integrated gate electrode. The differential conductance shows
sharp spikes corresponding to the addition of electrons to the
SWCNT along with side peaks due to phonon-assisted tunneling. The
entire Coulomb diamond plot is accessed using the gate electrode.
Furthermore, the gate electrode is used to confirm that the side
peaks arising at high currents correspond to excitations of the
main Coulomb peaks.  This supports the interpretation that they
represent phonon-assisted electron transport.

Our measurements leave three main unresolved issues that require
more theoretical and experimental work; 1) many of the SWCNTs show
Coulomb blockade behavior as prepared (i.e. without being cut by
the STM tip) implying that there is a tunneling barrier between
the SWCNT and the Pt substrate. More work needs to be done to
determine if this barrier is formed by defects near the edge of
the trench, band bending, or some other mechanism. 2)
Phonon-assisted side peaks are only observed near the center of
the suspended SWCNTs\cite{LeRoyNature}. Why are the phonons
preferentially excited in this region when the electronic
wavefunction is extended over the entire suspended region of the
SWCNT? 3) Previously, we have studied the strength of the
phonon-assisted side peaks as a function of current through the
SWCNT\cite{LeRoyNature}.  We found that the amplitude of the side
peaks scaled as a Bessel function in analogy to photon-assisted
tunneling\cite{Tien}. Experimentally we found the argument of the
Bessel function scaled linearly with the current. However, the
electron-phonon term in the Hamiltonian scales with the square
root of the number of phonons and hence current.  The origin of
this discrepancy must still be addressed with more experimental
measurements and theoretical work.

% Surround figure environment with turnpage environment for landscape
% figure
% \begin{figure}
% \end{figure}

%\appendix
%\section{}

\begin{acknowledgments}
The authors would like to thank Ya. Blanter and Y. Nazarov for
helpful discussions and I. Heller for help with the sample growth.
This research was supported by Stichting voor Fundamental
Onderzoek der Materie (FOM) and the Netherlands Organization for
Scientific Research (NWO).
\end{acknowledgments}

%\bibliography{LeRoyPRB}

\end{document}